\documentclass[twocolumn,showpacs,superscriptaddress,amsmath,amssymb,prl]{revtex4-2}
\usepackage[colorlinks,linkcolor=blue,anchorcolor=blue,citecolor=blue,urlcolor=blue]{hyperref}
\usepackage{graphicx}
\usepackage{epsfig}
\usepackage{dcolumn}
\usepackage{bm}
\usepackage{overpic}
\usepackage{color}
\usepackage{multirow}
\usepackage{subfigure}
\usepackage{lineno}
\usepackage{notoccite}

\usepackage{verbatim}
\usepackage{cleveref}

\begin{document}
\newcommand{\ks}{K_{S}^{0}}
\newcommand{\EP}{e^{+}}
\newcommand{\EM}{e^{-}}
\newcommand{\epm}{e^{\pm}}
\newcommand{\vpho}{\gamma^{\ast}}
\newcommand{\qqbar}{q\bar{q}}

\newcommand{\ee}{e^{+}e^{-}}
\newcommand{\mm}{\mu^{+}\mu^{-}}
\newcommand{\alfs}{\alpha_{s}}
\newcommand{\alfmz}{\alpha(M_{Z}^{2})}
\newcommand{\amu}{a_{\mu}}
\newcommand{\Lam}{\Lambda_{c}}
\newcommand{\lam}{\Lambda_{c}^{+}}
\newcommand{\lambar}{\bar{\Lambda}_{c}^{-}}
\newcommand{\Lambdac}{\Lambda_{c}}
\newcommand{\mbc}{M_{BC}}
\newcommand{\dele}{\Delta E}
\newcommand{\ebm}{E_{\textmd{beam}}}
\newcommand{\ecm}{E_{\textmd{c.m.}}}
\newcommand{\pbm}{p_{\textmd{beam}}}
\newcommand{\MuMu}{\mu\mu}
\newcommand{\mumu}{\mu\mu}
\newcommand{\tata}{\tau^{+}\tau^{-}}
\newcommand{\pipi}{\pi^{+}\pi^{-}}
\newcommand{\gaga}{\gamma\gamma}
\newcommand{\twopho}{\ee+X}
\newcommand{\sqs}{\sqrt{s}}
\newcommand{\sqsp}{\sqrt{s^{\prime}}}
\newcommand{\da}{\Delta\alpha}
\newcommand{\das}{\Delta\alpha(s)}
\newcommand{\dimu}{\ee \ra \mumu}
\newcommand{\dedx}{\textmd{d}E/\textmd{d}x}
\newcommand{\chip}{\chi_{\textmd{Prob}}}
\newcommand{\chiP}{\chi_{p}}
\newcommand{\evz}{V_{z}^{\textmd{evt}}}
\newcommand{\evzloose}{V_{z,\textmd{loose}}^{\textmd{evt}}}
\newcommand{\avz}{V_{z}^{\textmd{ave}}}
\newcommand{\Ngd}{N_{\textmd{good}}}
\newcommand{\Ncru}{N_{\textmd{crude}}}
\newcommand{\pio}{\pi^{0}}
\newcommand{\rpid}{r_{\textmd{PID}}}

\newcommand{\Nhxobs}{N_{h+X}^{\textmd{obs}}}
\newcommand{\Nhobs}{N_{h}^{\textmd{obs}}}
\newcommand{\Npioxobs}{N_{\pi^{0}+X}^{\textmd{obs}}}
\newcommand{\Nksxobs}{N_{\ks+X}^{\textmd{obs}}}
\newcommand{\Npioobs}{N_{\pi^{0}}^{\textmd{obs}}}
\newcommand{\Nksobs}{N_{\ks}^{\textmd{obs}}}
\newcommand{\Nhxtru}{N_{h+X}^{\textmd{tru}}}
\newcommand{\Nhtru}{N_{h}^{\textmd{tru}}}
\newcommand{\Npiotru}{N_{\pi^{0}}^{\textmd{tru}}}
\newcommand{\Nkstru}{N_{\ks}^{\textmd{tru}}}
\newcommand{\Nbarhxobs}{\bar{N}_{h+X}^{\textmd{obs}}}
\newcommand{\Nbarhobs}{\bar{N}_{h}^{\textmd{obs}}}
\newcommand{\Nbarpioobs}{\bar{N}_{\pi^{0}}^{\textmd{obs}}}
\newcommand{\Nbarksobs}{\bar{N}_{\ks}^{\textmd{obs}}}
\newcommand{\Nbarhxtru}{\bar{N}_{h+X}^{\textmd{tru}}}
\newcommand{\Nbarhtru}{\bar{N}_{h}^{\textmd{tru}}}
\newcommand{\Nbarpiotru}{\bar{N}_{\pi^{0}}^{\textmd{tru}}}
\newcommand{\Nbarkstru}{\bar{N}_{\ks}^{\textmd{tru}}}

\newcommand{\Nhadtot}{N_{\textmd{had}}^{\textmd{tot}}}
\newcommand{\Nhadobs}{N_{\textmd{had}}^{\textmd{obs}}}
\newcommand{\Nbarhadobs}{\bar{N}_{\textmd{had}}^{\textmd{obs}}}
\newcommand{\Nhadtru}{N_{\textmd{had}}^{\textmd{tru}}}
\newcommand{\Nbarhadtru}{\bar{N}_{\textmd{had}}^{\textmd{tru}}}
\newcommand{\Nhadphy}{N_{\textmd{had}}}

\newcommand{\cshadobs}{\sigma_{\textmd{had}}^{\textmd{obs}}}
\newcommand{\effhad}{\vap_{\textmd{had}}}
\newcommand{\efftrg}{\vap_{\textmd{trig}}}
\newcommand{\lint}{\mathcal{L}_{\textmd{int}}}
\newcommand{\Nbkg}{N_{\textmd{bkg}}}
\newcommand{\NbkgTot}{N_{\textrm{bkg}}^{\textrm{Tot}}}
\newcommand{\csbkg}{\sigma_{\textmd{bkg}}}
\newcommand{\Nmcsur}{N_{\textmd{MC}}^{\textmd{sur}}}
\newcommand{\Nmcsurori}{N_{\textmd{MC}}^{\textmd{sur,nom.}}}
\newcommand{\Nmcsurwtd}{N_{\textmd{MC}}^{\textmd{sur,wtd.}}}
\newcommand{\Nmcgen}{N_{\textmd{MC}}^{\textmd{gen}}}
\newcommand{\vap}{\varepsilon}
\newcommand{\chisq}{\chi^{2}}
\newcommand{\cshadphy}{\sigma_{\textmd{had}}^{\textmd{phy}}}
\newcommand{\cshadtot}{\sigma_{\textmd{had}}^{\textmd{tot}}}
\newcommand{\cshadborn}{\sigma_{\textmd{had}}^{0}}
\newcommand{\cshadborncon}{\sigma_{\textmd{con}}^{0}}
\newcommand{\cshadbornres}{\sigma_{\textmd{res}}^{0}}
\newcommand{\csdimuborn}{\sigma_{\mu\mu}^{0}}
\newcommand{\rpqcd}{R_{\textmd{pQCD}}}
\newcommand{\Nprod}{N_{\textmd{prod}}}
\newcommand{\Nhadnet}{N_{\textrm{had}}^{\textrm{net}}}
\newcommand{\Delrel}{\Delta_{\textrm{rel}}}

\newcommand{\fourpionchg}{\pipi\pipi}
\newcommand{\fourpionneu}{\pipi\pi^{0}\pi^{0}}
\newcommand{\sixpionchg}{3(\pipi)}
\newcommand{\thrpionneu}{\pipi\pi^{0}}
\newcommand{\twopionchg}{\pipi}

\newcommand{\Nsurnpion}{N_{\textmd{sur}}^{n\pi}}
\newcommand{\Ngennpion}{N_{\textmd{gen}}^{n\pi}}
\newcommand{\Ngentot}{N_{\textmd{gen}}^{\textmd{tot}}}
\newcommand{\effincnpion}{\vap_{n\pi}^{\textmd{inc}}}
\newcommand{\effincnpionp}{\vap_{n\pi}^{\textmd{inc},\prime}}
\newcommand{\effincnonnpion}{\vap_{\textmd{non}-n\pi}^{\textmd{inc}}}
\newcommand{\effexcnpion}{\vap_{n\pi}^{\textmd{exc}}}
\newcommand{\fracnpion}{f_{n\pi}}
\newcommand{\fracnpionp}{f_{n\pi}^{\prime}}
\newcommand{\fracnonnpion}{f_{\textmd{non}-n\pi}}

\newcommand{\Nsurtwopion}{N_{\textmd{sur}}^{2\pi}}
\newcommand{\Ngentwopion}{N_{\textmd{gen}}^{2\pi}}
\newcommand{\effinctwopion}{\vap_{2\pi}^{\textmd{inc}}}
\newcommand{\effinctwopionp}{\vap_{2\pi}^{\textmd{inc},\prime}}
\newcommand{\effincnontwopion}{\vap_{\textmd{non}-2\pi}^{\textmd{inc}}}
\newcommand{\effexctwopion}{\vap_{2\pi}^{\textmd{exc}}}
\newcommand{\fractwopion}{f_{2\pi}}
\newcommand{\fractwopionp}{f_{2\pi}^{\prime}}
\newcommand{\fracnontwopion}{f_{\textmd{non}-2\pi}}

\newcommand{\Nsurthrpion}{N_{\textmd{sur}}^{3\pi}}
\newcommand{\Ngenthrpion}{N_{\textmd{gen}}^{3\pi}}
\newcommand{\effincthrpion}{\vap_{3\pi}^{\textmd{inc}}}
\newcommand{\effincthrpionp}{\vap_{3\pi}^{\textmd{inc},\prime}}
\newcommand{\effincnonthrpion}{\vap_{\textmd{non}-3\pi}^{\textmd{inc}}}
\newcommand{\effexcthrpion}{\vap_{3\pi}^{\textmd{exc}}}
\newcommand{\fracthrpion}{f_{3\pi}}
\newcommand{\fracthrpionp}{f_{3\pi}^{\prime}}
\newcommand{\fracnonthrpion}{f_{\textmd{non}-3\pi}}

\newcommand{\Nsurfourpion}{N_{\textmd{sur}}^{4\pi}}
\newcommand{\Ngenfourpion}{N_{\textmd{gen}}^{4\pi}}
\newcommand{\effincfourpion}{\vap_{4\pi}^{\textmd{inc}}}
\newcommand{\effincfourpionp}{\vap_{4\pi}^{\textmd{inc},\prime}}
\newcommand{\effincnonfourpion}{\vap_{\textmd{non}-4\pi}^{\textmd{inc}}}
\newcommand{\effexcfourpion}{\vap_{4\pi}^{\textmd{exc}}}
\newcommand{\fracfourpion}{f_{4\pi}}
\newcommand{\fracfourpionp}{f_{4\pi}^{\prime}}
\newcommand{\fracnonfourpion}{f_{\textmd{non}-4\pi}}

\newcommand{\Npionprod}{N_{\textmd{prod}}^{4\pi}}
\newcommand{\Ndatasur}{N_{\textmd{data}}^{\textmd{sur}}}
\newcommand{\Nobspion}{N_{\textmd{obs}}^{4\pi}}
\newcommand{\Nhadprod}{N_{\textmd{prod}}^{\textmd{had}}}
\newcommand{\sigmaobs}{\sigma_{\textmd{obs}}}
\newcommand{\effhadp}{\vap_{\textmd{had}}^{\prime}}

\newcommand{\effpion}{\vap_{4\pi}}
\newcommand{\effexcpion}{\vap_{4\pi}^{\textmd{exc}}}
\newcommand{\effincpion}{\vap_{4\pi}^{\textmd{inc}}}
\newcommand{\effincpionI}{\vap_{4\pi}^{\textmd{inc},1}}
\newcommand{\effincpionII}{\vap_{4\pi}^{\textmd{inc},2}}
\newcommand{\effincpionp}{\vap_{4\pi}^{\textmd{inc},\prime}}
\newcommand{\effincremain}{\vap_{\textmd{non}-n\pi}^{\textmd{inc}}}

\newcommand{\fracpion}{f_{4\pi}}
\newcommand{\fracnonpion}{f_{\textmd{non}-4\pi}}
\newcommand{\fracnonpionp}{f_{\textmd{non}-4\pi}^{\prime}}
\newcommand{\fracpionII}{f_{4\pi}^{2}}
\newcommand{\fracpionp}{f_{4\pi}^{\prime}}
\newcommand{\reladiff}{\Delta_{\textmd{rel}}}

\newcommand{\Nsursixpion}{N_{\textmd{sur}}^{6\pi}}
\newcommand{\Ngensixpion}{N_{\textmd{gen}}^{6\pi}}
\newcommand{\effincsixpion}{\vap_{6\pi}^{\textmd{inc}}}
\newcommand{\fracsixpion}{f_{6\pi}}

\newcommand{\etot}{E_{\textmd{tot}}}
\newcommand{\ptot}{p_{\textmd{tot}}}
\newcommand{\plab}{p_{\textmd{Lab}}}
\newcommand{\mpiOI}{M(\pi^{0}_{1})}
\newcommand{\mpiOII}{M(\pi^{0}_{2})}

\newcommand{\widtheeoi}{\varGamma^{\textmd{ee}}_{0,i}}
\newcommand{\widtheeoj}{\varGamma^{\textmd{ee}}_{0,j}}
\newcommand{\widtheeo}{\varGamma^{\textmd{ee}}_{0}}
\newcommand{\widthee}{\varGamma^{\textmd{ee}}}
\newcommand{\widtheeexpi}{\varGamma^{\textmd{ee}}_{\textmd{exp},i}}
\newcommand{\widtheeexp}{\varGamma^{\textmd{ee}}_{\textmd{exp}}}
\newcommand{\widthtoti}{\varGamma^{\textmd{tot}}_{i}}
\newcommand{\widthtot}{\varGamma^{\textmd{tot}}}

\newcommand{\vpqed}{\Pi_{\textmd{QED}}}
\newcommand{\vpqcd}{\Pi_{\textmd{QCD}}}
\newcommand{\vpcon}{\Pi_{\textmd{con}}}
\newcommand{\vpres}{\Pi_{\textmd{res}}}
\newcommand{\vpo}{\Pi_{0}}
\newcommand{\rcon}{R_{\textmd{con}}}
\newcommand{\rres}{R_{\textmd{res}}}
\newcommand{\rexp}{R_{\textmd{exp}}}

\newcommand{\delvert}{\delta_{\textmd{vert}}}
\newcommand{\delvp}{\delta_{\textmd{vac}}}
\newcommand{\delbrem}{\delta_{\gamma}}
\newcommand{\delobs}{\delta_{\textmd{obs}}}
\newcommand{\radiatorsf}{F_{\textmd{SF}}}
\newcommand{\radiatorfd}{F_{\textmd{FD}}}
\newcommand{\DelFD}{\Delta_{\textmd{FD}}}
\newcommand{\DelFDCal}{\Delta_{\textmd{cal}}}
\newcommand{\DelFDcs}{\Delta_{\sigma}}
\newcommand{\DelFDvp}{\Delta_{\textmd{vp}}}

\newcommand{\costh}{\cos\theta}
\newcommand{\costhIprg}{\cos\theta_{\textmd{1prg}}}
\newcommand{\costhIIprg}{\cos\theta_{\textmd{2prg}}}
\newcommand{\costhIIIprg}{\cos\theta_{\textmd{3prg}}}
\newcommand{\costhIVprg}{\cos\theta_{\textmd{4prg}}}
\newcommand{\costhrestprg}{\cos\theta_{\textmd{restprg}}}
\newcommand{\emce}{E^{\textmd{ctrk.}}_{\textmd{emc}}}
\newcommand{\emceIprg}{E^{\textmd{ctrk.}}_{\textmd{emc,1prg}}}
\newcommand{\emceIIprg}{E^{\textmd{ctrk.}}_{\textmd{emc,2prg}}}
\newcommand{\emceIIIprg}{E^{\textmd{ctrk.}}_{\textmd{emc,3prg}}}
\newcommand{\emceIVprg}{E^{\textmd{ctrk.}}_{\textmd{emc,4prg}}}
\newcommand{\emcerestprg}{E^{\textmd{ctrk.}}_{\textmd{emc,restprg}}}
\newcommand{\isocosth}{\cos\theta_{\textmd{iso}}}
\newcommand{\isocosthIprg}{\cos\theta_{\textmd{iso,1prg}}}
\newcommand{\isocosthIIprg}{\cos\theta_{\textmd{iso,2prg}}}
\newcommand{\isocosthIIIprg}{\cos\theta_{\textmd{iso,3prg}}}
\newcommand{\isocosthIVprg}{\cos\theta_{\textmd{iso,4prg}}}
\newcommand{\isocosthrestprg}{\cos\theta_{\textmd{iso,restprg}}}
\newcommand{\eop}{E/P}
\newcommand{\eopIprg}{E/P_{\textmd{1prg}}}
\newcommand{\eopIIprg}{E/P_{\textmd{2prg}}}
\newcommand{\eopIIIprg}{E/P_{\textmd{3prg}}}
\newcommand{\eopIVprg}{E/P_{\textmd{4prg}}}
\newcommand{\eoprestprg}{E/P_{\textmd{restprg}}}
\newcommand{\nisogam}{N_{\textmd{isogam}}}
\newcommand{\nisogamIprg}{N_{\textmd{isogam,1prg}}}
\newcommand{\nisogamIIprg}{N_{\textmd{isogam,2prg}}}
\newcommand{\nisogamIIIprg}{N_{\textmd{isogam,3prg}}}
\newcommand{\nisogamIVprg}{N_{\textmd{isogam,4prg}}}
\newcommand{\nisogamrestprg}{N_{\textmd{isogam,restprg}}}
\newcommand{\ptrk}{p_{\textmd{ctrk}}}
\newcommand{\pIprg}{p^{\textmd{ctrk}}_{\textmd{1prg}}}
\newcommand{\pIIprg}{p^{\textmd{ctrk}}_{\textmd{2prg}}}
\newcommand{\pIIIprg}{p^{\textmd{ctrk}}_{\textmd{3prg}}}
\newcommand{\pIVprg}{p^{\textmd{ctrk}}_{\textmd{4prg}}}
\newcommand{\prestprg}{p^{\textmd{ctrk}}_{\textmd{restprg}}}
\newcommand{\tote}{E_{\textmd{vis.}}}
\newcommand{\totevte}{E_{\textmd{tot.}}}
\newcommand{\totevteIprg}{E_{\textmd{tot.}}^{\textmd{1prg}}}
\newcommand{\balanceIprg}{\textmd{Balance}}
\newcommand{\ngamma}{N_{\gamma}}
\newcommand{\ngammaIprg}{N_{\gamma,\textmd{1prg}}}
\newcommand{\ngammaIIprg}{N_{\gamma,\textmd{2prg}}}
\newcommand{\ngammaIIIprg}{N_{\gamma,\textmd{3prg}}}
\newcommand{\ngammaIVprg}{N_{\gamma,\textmd{4prg}}}
\newcommand{\ngammarestprg}{N_{\gamma,\textmd{restprg}}}
\newcommand{\ngoodwt}{N_{\textmd{good}}^{\textmd{Wt}}}
\newcommand{\ngood}{N_{\textmd{good}}}
\newcommand{\npiO}{N_{\pi^{0}}}
\newcommand{\npiOIprg}{N_{\pi^{0}}^{\textmd{1prg}}}
\newcommand{\npiOIIprg}{N_{\pi^{0}}^{\textmd{2prg}}}
\newcommand{\npp}{N_{p}}
\newcommand{\npm}{N_{\bar{p}}}
\newcommand{\nkp}{N_{K^{+}}}
\newcommand{\nkm}{N_{K^{-}}}
\newcommand{\npip}{N_{\pi^{+}}}
\newcommand{\npim}{N_{\pi^{-}}}
\newcommand{\ppp}{P(p^{+})}
\newcommand{\ppm}{P(\bar{p}^{-})}
\newcommand{\pkp}{p(K^{+})}
\newcommand{\pkm}{p(K^{-})}
\newcommand{\ppip}{P(\pi^{+})}
\newcommand{\ppim}{P(\pi^{-})}
\newcommand{\ppiO}{P(\pi^{0})}
\newcommand{\mpiO}{M(\pi^{0})}
\newcommand{\mks}{M(K^{0}_{s})}
\newcommand{\pks}{p_{K^{0}_{s}}}
\newcommand{\mphi}{M(\phi)}
\newcommand{\pphi}{p_{\phi}}
\newcommand{\mIIgam}{M(\gamma\gamma)}
\newcommand{\mIIgamIprg}{M(\gamma\gamma)^{\textmd{1prg}}}
\newcommand{\pIIgam}{p_{\gamma\gamma}}
\newcommand{\mlambda}{M(\Lambda)}
\newcommand{\plambda}{p_{\Lambda}}
\newcommand{\mdO}{M(D^{0})}
\newcommand{\pdO}{p_{D^{0}}}
\newcommand{\mdstarO}{M(D^{\ast 0})}
\newcommand{\pdstarO}{p_{D^{\ast 0}}}
\newcommand{\mdp}{M(D^{\pm})}
\newcommand{\pdp}{p_{D^{\pm}}}
\newcommand{\mdstarp}{M(D^{\ast\pm})}
\newcommand{\pdstarp}{p_{D^{\ast\pm}}}
\newcommand{\mds}{M(D_{s}^{\pm})}
\newcommand{\pds}{p_{D_{s}^{\pm}}}
\newcommand{\mdstars}{M(D_{s}^{\ast\pm})}
\newcommand{\pdstars}{p_{D_{s}^{\ast\pm}}}
\newcommand{\Vr}{V_{r}}
\newcommand{\Vz}{V_{z}}

\newcommand{\gev}{\mathrm{GeV}}
\newcommand{\mev}{\mathrm{MeV}}
\newcommand{\mevcc}{\mathrm{MeV}/c^{2}}
\newcommand{\gevc}{\mathrm{GeV}/c}
\newcommand{\gevcc}{\mathrm{GeV}/c^2}

\newcommand{\nchg}{N_{\textmd{chg}}}
\newcommand{\eff}{\vap}

\newcommand{\critecm}{1.780}

\newcommand{\ENERGYAT}{4575.5}
\newcommand{\ENERGYBT}{4575.5}
\newcommand{\ENERGYCT}{4575.5}
\newcommand{\ENERGYDT}{4575.5}
\newcommand{\ksdecay}{\ks\ra\pi^{+}\pi^{-}}
\newcommand{\phidecay}{\phi\ra K^{+}K^{-}}
\newcommand{\piOdecay}{\pi^{0}\ra\gamma\gamma}
\newcommand{\Lambdadecay}{\Lambda\ra p\pi^{-}}
\newcommand{\DOdecay}{D^{0}\ra K^{-}\pi^{+}}
\newcommand{\DStarOdecay}{D^{\ast0}\ra D^{0}\pi^{0}}
\newcommand{\Dpdecay}{D^{+}\ra K^{+}\pi^{+}\pi^{-}}
\newcommand{\DStarpdecay}{D^{\ast+}\ra D^{0}\pi^{+}}
\newcommand{\Dsdecay}{D^{+}_{s}\ra K^{+}K^{-}\pi^{+}}
\newcommand{\DStarsdecay}{D^{\ast+}_{s}\ra D^{+}_{s}\gamma}


\title{\boldmath \textbf{Single Inclusive $\pi^\pm$ and $K^\pm$ Production in $e^+e^-$ Annihilation \\at center-of-mass Energies from 2.000 to 3.671~GeV}}

\author{
M.~Ablikim$^{1}$, M.~N.~Achasov$^{4,c}$, P.~Adlarson$^{77}$, X.~C.~Ai$^{82}$, R.~Aliberti$^{36}$, A.~Amoroso$^{76A,76C}$, Q.~An$^{73,59,a}$, Y.~Bai$^{58}$, O.~Bakina$^{37}$, Y.~Ban$^{47,h}$
% ...
\\
\vspace{0.2cm}
(BESIII Collaboration)\\
\vspace{0.2cm} {\it
$^{1}$ Institute of High Energy Physics, Beijing 100049, People's Republic of China\\
% ...
$^{4}$ Budker Institute of Nuclear Physics SB RAS (BINP), Novosibirsk 630090, Russia\\
% ...
$^{36}$ Johannes Gutenberg University of Mainz, Johann-Joachim-Becher-Weg 45, D-55099 Mainz, Germany\\
$^{37}$ Joint Institute for Nuclear Research, 141980 Dubna, Moscow region, Russia\\
% ...
$^{47}$ Peking University, Beijing 100871, People's Republic of China\\
% ...
$^{58}$ Southeast University, Nanjing 211100, People's Republic of China\\
$^{59}$ State Key Laboratory of Particle Detection and Electronics, Beijing 100049, Hefei 230026, People's Republic of China\\
% ...
$^{73}$ University of Science and Technology of China, Hefei 230026, People's Republic of China\\
% ...
$^{76}$ University of Turin and INFN, (A)University of Turin, I-10125, Turin, Italy; (B)University of Eastern Piedmont, I-15121, Alessandria, Italy; (C)INFN, I-10125, Turin, Italy\\
$^{77}$ Uppsala University, Box 516, SE-75120 Uppsala, Sweden\\
% ...
$^{82}$ Zhengzhou University, Zhengzhou 450001, People's Republic of China\\
\vspace{0.2cm}
$^{a}$ Deceased\\
% ...
$^{c}$ Also at the Novosibirsk State University, Novosibirsk, 630090, Russia\\
% ...
$^{h}$ Also at State Key Laboratory of Nuclear Physics and Technology, Peking University, Beijing 100871, People's Republic of China\\
% ...
}
}


\begin{abstract}
Using data samples with a total integrated luminosity of 253~$\rm pb^{-1}$ collected by the BESIII detector operating at the BEPCII collider,
the differential cross sections of inclusive $\pi^\pm$ and $K^\pm$ production, as a function of momentum and normalized by the total hadronic cross section, are measured at center-of-mass energies from 2.000 to 3.671~GeV.
The measured $\pi^{\pm}$ cross sections are consistent with the previously reported $\pi^{0}$ cross sections by BESIII, while the $K^{\pm}$ cross sections are systematically higher than the $K^0_S$ cross sections by a factor of approximately 1.4.
These new results are in agreement with state-of-the-art QCD analyses at next-to-next-to-leading-order accuracy, particularly in the large hadron momentum region at energy scales down to 3~GeV.
These findings support the validity of isospin symmetry in parton fragmentation processes.

\end{abstract}

\maketitle


Single inclusive hadron production in electron-positron annihilation (SIA) provides direct insights into the hadronization process in a controlled and clean environment. It enhances our understanding of how quarks and gluons fragment into hadrons, which is crucial for advancing our knowledge of quantum chromodynamics (QCD) and for accurately modeling hadron production in various high-energy processes.  The typical experimental observable
in SIA is
\begin{equation}{\label{eq:obs1}}
    \frac{1}{\sigma(e^+e^-\to \rm{hadrons})}\frac{d\sigma(e^+e^-\to h+X)}{dp_h},
\end{equation}
where $\sigma(e^+e^-\to \rm{hadrons})$ is the cross section for $e^+e^-$ annihilation to all possible hadronic final states (referred to as inclusive hadronic events hereafter), $p_h$ represents the momentum of the identified hadron $h$, $X$ refers to everything else.
This observable in high-energy collisions can be factorized using the QCD factorization theorem~\cite{Collins:1989gx}, which allows
it to be expressed as a convolution of perturbative hard-part coefficients
and nonperturbative fragmentation functions (FFs).
At leading order in $\alpha_{s}$, this observable can be interpreted as $\Sigma_q e_q^2 [D_q^h(z, \mu)+D_{\bar{q}}^h(z, \mu)]$, where $e_{q}$ is the fractional charge of the quark $q$, and $D^h_{q/\bar{q}}(z,\mu)$ is the FF presenting the probability density that an outgoing parton (quark $q$ or antiquark $\bar{q}$)
produces a hadron $h$. The parameter $\mu$ is the factorization scale, typically chosen to be
the center-of-mass (c.m.) energy $\sqrt{s}$.
The dimensionless variable $z\equiv2\sqrt{p^2_{h}c^2+M^2_{h}c^4}/\sqrt{s}$ denotes the relative energy of hadron $h$ with mass $M_h$.

In $e^+e^-$ collisions, a broad range of measurements is available for the single inclusive production of identified light charged hadrons, including $\pi^{\pm}$,
$K^{\pm}$, $p/\bar p$, as well as unidentified charged hadrons, which have been summarized comprehensively, e.g.~in Table IV of Ref.~\cite{Gao:2024dbv}. These measurements are primarily concentrated in the higher-energy region. The Belle~\cite{Belle:2013lfg,Belle:2024vua} and BARBAR~\cite{BaBar:2013yrg} Collaborations have performed precision measurements at around 10.5~GeV.
In the energy region below 10~GeV, the only available measurements of inclusive $\pi^{\pm}/K^{\pm}$ production come from the DASP experiment~\cite{DASP:1978ftr}, conducted in 1978, mainly focusing on the contribution of charm to charged hadron production near the charm threshold.
However, there is a gap in measurements at lower-energy scales, especially in the continuum region $2-3$~GeV.
In this Letter, we present a study of the processes \mbox{$e^{+}e^{-}\to \pi^{\pm}/K^{\pm} + X$} using datasets collected at eight c.m.~energies from 2.000 to 3.671~GeV, with $z$ coverage from 0.13 to 0.95 and from 0.30 to 0.95 for $\pi^\pm$ and $K^{\pm}$, respectively.
The results provide a unique opportunity to test QCD factorization at low-energy scales and to
assess the consistency of charged hadron production between
$e^+e^-$ collisions and semi-inclusive deep inelastic scattering  measurements from COMPASS~\cite{Alexeev:2024krc}, HERMES~\cite{HERMES:2012uyd}, and Jefferson Lab~\cite{JeffersonLabHallA:2014yxb}.
Furthermore, with a variety of identified final-state hadrons,
it allows for the exploration of intriguing QCD phenomena, such as testing isospin symmetry.
In particular, recent comparisons of the yields of $K_S^0$ and $K^{\pm}$ in high-energy nuclear
collisions suggest potential violations of isospin symmetry, which contradicts the conclusions derived
from mass measurements of these particles~\cite{NA61SHINE:2023azp}.
The SIA process, as the cleanest process to detect particle productions, provides a unique and precise probe for investigating such fundamental effects and clarifying the role of isospin symmetry in the parton fragmentation processes.

The datasets used in this Letter were collected with the
BESIII detector~\cite{BESIII:2009fln} running at BEPCII~\cite{Yu:2016cof}.
Experimentally, the normalized differential cross section for the inclusive production of the identified $\pi^{\pm}/K^{\pm}$ in Eq.~(\ref{eq:obs1}) is determined with
\begin{equation}{\label{eq:exobs1d}}
    \frac{N^{\rm{obs}}_{\pi^{\pm}/K^{\pm}}}{N^{\rm{obs}}_{\rm{had}}}\frac{1}{\Delta p}f_{\pi^{\pm}/K^{\pm}},
\end{equation}
where $N^{\rm{obs}}_{\pi^{\pm}/K^{\pm}}$ is the number of $e^+e^-\to \pi^{\pm}/K^{\pm}+X$ events within a certain momentum range $\Delta p$ (referred to as the momentum bin hereafter), $N^{\rm obs}_{\rm had}$ represents the number of observed hadronic events in the $e^+e^-$ annihilation at a given c.m.~energy, and $f_{\pi^{\pm}/K^{\pm}}$ is the correction factor accounting for the detection efficiency and initial-state radiation effects.

To identify the signal events, the inclusive hadronic events are first selected~\cite{BESIII:2021wib}. The dominant background processes, $e^+e^-\to e^+e^-$ (Bhabha scattering) and $e^+e^-\to\gamma\gamma$, are rejected by applying dedicated requirements on the showers recorded by the electromagnetic calorimeter.
In the remaining events, a set of selection criteria are implemented to identify the good charged tracks (prongs). Events with zero or one prong are removed to suppress the contribution of quantum electrodynamics-(QED) related and beam-associated backgrounds.
For events with two or three prongs, further requirements are employed to suppress QED-related backgrounds.
Events with more than three prongs are regarded as hadronic events directly.
More details of the selection of inclusive hadronic events are described in Ref.~\cite{BESIII:2021wib}.

Although comprehensive selection criteria have been used to select the inclusive hadronic events, residual background still exists in the data.
The yields of residual QED-related background events are estimated by analyzing the corresponding Monte Carlo (MC) simulation samples, which are produced based on
the {\sc geant4} software~\cite{GEANT4:2002zbu}.
In these simulations, the geometric description of the BESIII detector and its interaction with particles are implemented.
The background processes $e^+e^-\to e^+e^-$, $\mu^+\mu^-$, and $\gamma\gamma$ are generated by the {\sc babayaga3.5} tool~\cite{CarloniCalame:2000pz}.
At $\sqrt{s}=3.671$~GeV, which is above the threshold of $\tau^+\tau^-$ pair production, the $e^+e^-\to\tau^+\tau^-$ process is simulated with the {\sc kkmc} program~\cite{Jadach:1999vf} and the decay of the $\tau$ lepton is modeled by {\sc evtgen}~\cite{Lange:2001uf,Ping:2008zz}.
The two-photon processes $e^+e^-\to e^+e^- X$ with $X=e^+e^-(\mu^+\mu^-)$, $\eta(\eta')$, and $\pi^+\pi^-(K^+K^-)$ are simulated using the generators {\sc diag36}~\cite{Berends:1986ig}, {\sc ekhara}~\cite{Czyz:2017veo}, and {\sc galuga2.0}~\cite{Schuler:1997ex}, respectively.
In addition, the beam-associated background is estimated with a sideband method developed in Ref.~\cite{BESIII:2021wib}. A variable $V_z^{\rm evt}$, representing the average vertex position along the beam direction
of all charged tracks, is used to estimate the beam-associated background, with events in the sideband region of $V_z^{\rm evt}$ assumed to be beam associated.

Table~\ref{data_info} summarizes the integrated luminosities, the number of total selected inclusive hadronic events $(N^{\rm tot}_{\rm had})$ and the total remaining background events $(N_{\rm bkg})$ at each c.m.~energy, where $N^{\rm obs}_{\rm had}=N^{\rm tot}_{\rm had}-N_{\rm bkg}$.
\begin{table}[h!]
\begin{center}
\caption{The integrated luminosities and the numbers of total selected hadronic and residual background events in different c.m.~energies.}
\label{data_info}
\begin{tabular*}{\hsize}{@{}@{\extracolsep{\fill}}cccc@{}}\hline  \hline
$\sqrt{s}$~(GeV)      &$\mathcal{L}$~(pb$^{-1}$) &$N^{\rm tot}_{\rm had}$ &$N_{\rm bkg}$ \\ \hline
$2.0000$        &$10.074$   &$350 298\pm592$   &$8722\pm94$   \\
$2.2000$        &$13.699$   &$445 019\pm668$   &$10 737\pm104$ \\
$2.3960$        &$66.869$   &$1 869 906\pm1368$ &$47 550\pm219$ \\
$2.6444$        &$33.722$   &$817 528\pm905$   &$21 042\pm146$ \\
$2.9000$        &$105.253$  &$2 197 328\pm1483$ &$56 841\pm239$ \\
$3.0500$        &$14.893$   &$283 822\pm533$   &$7719\pm88$   \\
$3.5000$        &$3.633$    &$62 670\pm251$    &$1691\pm42$   \\
$3.6710$        &$4.628$    &$75 253\pm275$    &$6461\pm81$   \\ \hline \hline
\end{tabular*}
\end{center}
\end{table}

From the inclusive hadronic events, the $\pi^{\pm}/K^{\pm}$ mesons are selected with the particle identification (PID), which combines the measurements of the specific ionization energy loss
in the multilayer drift chamber (MDC) and the flight time from interaction point to the time-of-flight counter to form the probability $\mathcal{P}(h)~(h=\pi,~K,~p)$ under each hadron hypothesis.
The charged track is identified as the hadron species resulting in the highest probability.
Because of the misidentification effect, the raw counts ($N_{h^{\pm}}^{\rm raw}$), obtained after applying the PID requirements and subtracting the normalized numbers of remaining background tracks in each momentum bin, are written as
\begin{equation}
N_{h^{\pm}}^{\rm raw} = \sum_{g=\pi,K,p}\epsilon_{g^{\pm} \to h^{\pm}}N_{g^{\pm}}^{\rm obs},
\label{eq:Nraws}
\end{equation}
where $N_{g^{\pm}}^{\rm obs}$ is the number of hadron $g^{\pm}$ free of misidentification. The $\epsilon_{g^{\pm} \to h^{\pm}}$ terms stand for the particle identification ($g=h$) or misidentification ($g\neq h$) efficiencies, which constitute a PID efficiency matrix.
Accordingly, the transition from the raw counts to the observed counts requires the inversion of the PID efficiency matrix.
In our study, the contamination of electron or muon to $\pi/K$ sample is strongly suppressed due to the selection criteria of hadronic events and is therefore neglected. 
Nevertheless, these contributions are considered in the systematic uncertainty study.
In addition, the misidentifications from the oppositely charged hadrons are ignored due to the corresponding low efficiencies.

The PID efficiencies are studied using the $\pi^{\pm}$ control sample via $J/\psi\to\pi^+\pi^-\pi^0$, $K^{\pm}$ via $J/\psi\to K^0_S K^{\pm}\pi^{\mp}$, and the proton via $J/\psi, \psi(3686)\to p\bar{p}\pi^{+}\pi^{-}$.
To avoid possible bias introduced by the difference between the signal and control samples, the PID efficiencies are evaluated and the corresponding $N_{g^{\pm}}^{\rm obs}$ are extracted in each momentum versus ten bins in $\cos\theta$ bin $(p,\cos\theta)$, where $\theta$ is the polar angle with respect to the symmetry axis of MDC.
The contributions to $N_{\pi^{\pm}/K^{\pm}}^{\rm raw}$ from the residual QED-related and beam-associated backgrounds are subtracted.
The dominant background contribution is from Bhabha events, which is at most 10\% in a few bins~\cite{bhabhabkgs} and negligible in the other bins. The number of observed $\pi^{\pm}/K^{\pm}$ in each momentum bin is obtained by summing over all the $\cos\theta$ bins, i.e.,
\begin{equation}{\label{eq:Nobs}}
    N^{\rm obs}_{\pi^{\pm}/K^{\pm}}(i)=\sum_{j=1}^{10}N^{\rm obs}_{\pi^{\pm}/K^{\pm}}(i,j),
\end{equation}
where $N^{\rm obs}_{\pi^{\pm}/K^{\pm}}(i)$ is the number of observed $\pi^{\pm}/K^{\pm}$ in the $i$th momentum bin and $N^{\rm obs}_{\pi^{\pm}/K^{\pm}}(i,j)$ is that in the $i$th momentum and $j$th $\cos\theta$ bin.

\begin{figure*}[!htbp]
\begin{center}
\begin{overpic}[width=0.90\textwidth]{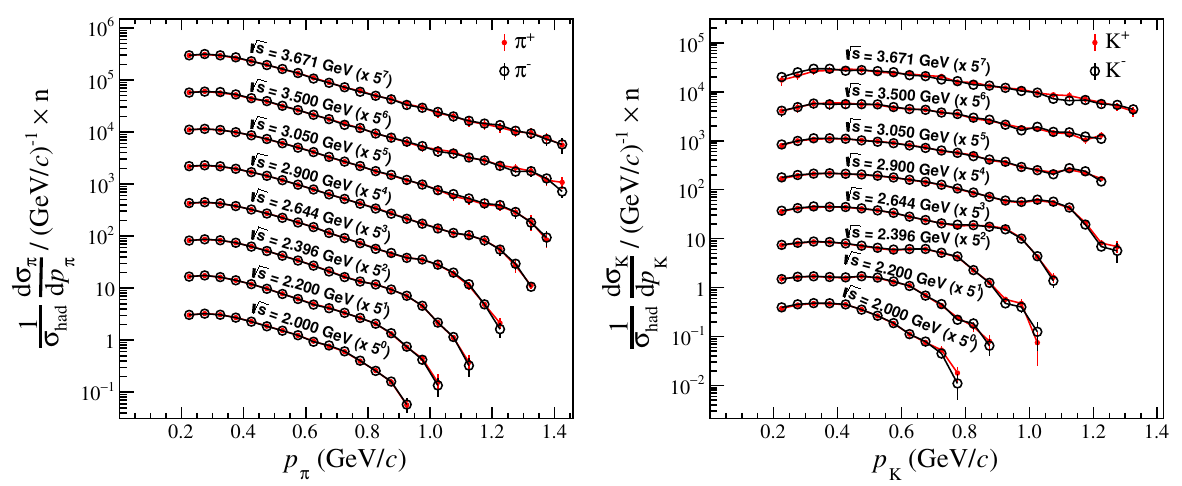}
\end{overpic}
\end{center}
\vspace*{-0.5cm}
\caption{Normalized differential cross sections of $e^+e^-\to\pi^{\pm}+X$ and $e^+e^-\to K^{\pm}+X$. The points with error bars are the measured
values, where the uncertainties are the quadrature sum of the corresponding statistical and systematic uncertainties. Note that the observable has been scaled by an artificial factor in the plots
and the line through the bin center is to group the points according to their center-of-mass energy.
The cross section of the process $e^+e^-\to K^{\pm}+X$ exhibits enhancements in the high-momentum regions. These enhancements may be attributed to contributions from the process $e^+e^-\to K_2^{*\pm}(1430)K^\mp, K_1^{\pm}(1400)K^{\mp}, K_1^{\pm}(1270)K^{\mp}$~\cite{BESIII:2022wxz,BESIII:2020vtu}.
}
\label{final_results_all}
\end{figure*}

The inclusive hadronic events are simulated with the {\sc luarlw} generator~\cite{Andersson:1997xwk,BESIII:2021wib,Andersson:1999ui}, in which, among others, the signal processes $e^+e^-\to\pi^{\pm}/K^{\pm}+X$ are included.
A detailed comparison between the MC sample and the experimental data
shows that the {\sc luarlw} model can reasonably describe the kinematic distributions of signal events in the data.
The correction factor $f_{\pi^{\pm}/K^{\pm}}$ can be written as
\begin{equation}
f_{\pi^{\pm}/K^{\pm}}=\frac{\bar{N}^{\rm tru}_{\pi^\pm/K^\pm} (\rm off)}{\Nbarhadtru(\rm off)}\bigg/\frac{\bar{N}^{\rm obs}_{\pi^\pm/K^\pm} (\rm on)}{\Nbarhadobs(\rm on)},
\end{equation}
where the variable $\bar{N}$ denotes the number of events determined from the inclusive hadronic MC sample, either after the detector reconstruction, similar to the experimental data, with superscript ``obs" or at truth level with superscript ``tru".
The terms ``on” and ``off” in the parentheses indicate that the corresponding quantities are extracted from the inclusive hadronic MC sample with or without simulating the initial-state radiation process, respectively.
To avoid misidentification in the determination of $N^{\rm obs}_{\pi^{\pm}/K^{\pm}}(\rm on)$, a match between the reconstructed $\pi^{\pm}/K^{\pm}$ candidates and truth-level $\pi^{\pm}/K^{\pm}$ mesons is performed instead of applying the PID on the MC sample.
The reconstructed $\pi^{\pm}/K^{\pm}$ candidate is matched with a truth-level $\pi^{\pm}/K^{\pm}$ meson if the opening angle (match angle) between their momenta is the smallest and less than $5^{\circ}$.

The sources of the systematic uncertainty in this analysis are categorized into two groups: those associated with the correction factor $f_{\pi^{\pm}/K^{\pm}}$, such as the MC model and the requirement of the match angle, and those related to the experimental observable $N^{\rm obs}_{\pi^{\pm}/K^{\pm}}/N^{\rm obs}_{\rm had}$, including the PID efficiency matrix, the misidentification from electrons and muons, and the hadronic event selection.

The dominant systematic uncertainty in this analysis is introduced by the MC simulation model of the inclusive hadronic events.
The generated fractions of the exclusive processes containing $\pi^{\pm}$ and $K^{\pm}$, which make up the signal processes $e^+e^-\to \pi^{\pm}/K^{\pm}+X$, directly affect the correction factors $f_{\pi^{\pm}}$ and $f_{K^{\pm}}$. To investigate the corresponding uncertainty, the {\sc hybrid} model~\cite{BESIII:2021wib,Rodrigo:2001kf,Ping:2016pms} is used as an alternative to simulate the inclusive hadronic events and reevaluate the correction factors.
In the {\sc hybrid} model, by taking the corresponding measured cross sections and production mechanisms into account, much knowledge of the allowed exclusive processes in the BESIII energy region is implemented.
The relative differences of the correction factors obtained with the nominal and alternative MC models are regarded as systematic uncertainties.
Higher uncertainties in the higher-momentum region are observed, which are due to the different production cross sections and intermediate kinematics of some few-body exclusive channels containing $\pi^{\pm}$ and $K^{\pm}$ between the two models.
The systematic uncertainty due to the match angle requirement is estimated by varying the acceptance threshold from $5^{\circ}$ to $10^{\circ}$.

\begin{figure*}[!htbp]
\begin{center}
\begin{overpic}[width=0.90\textwidth]{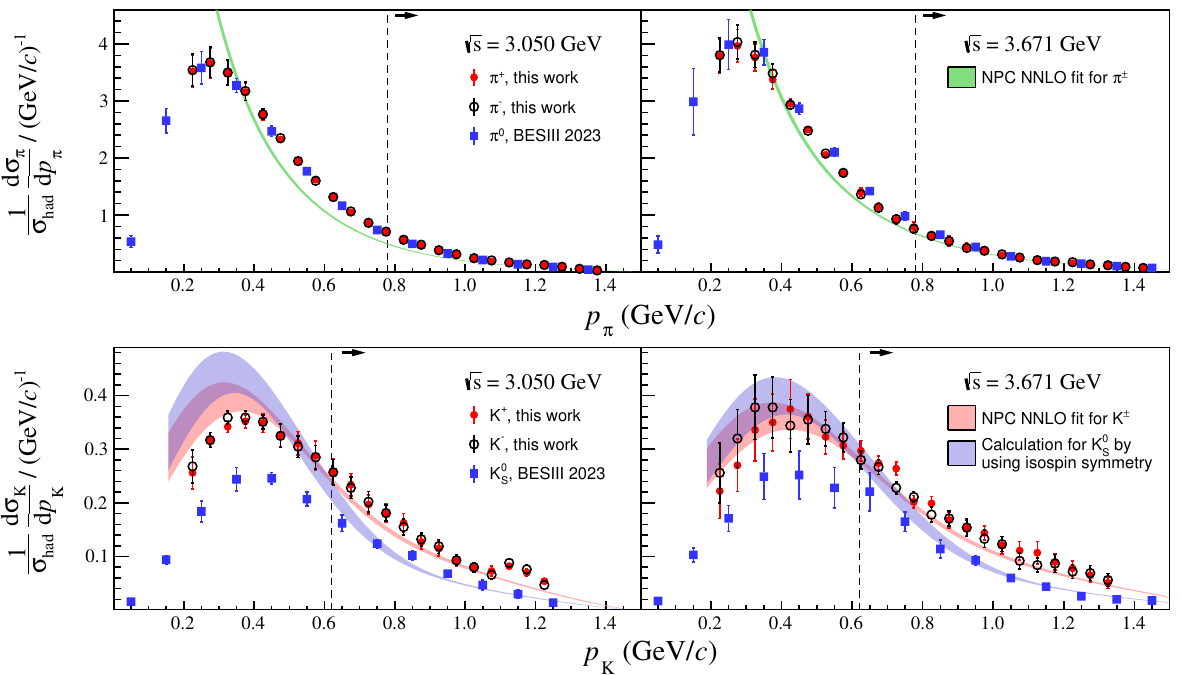}
\end{overpic}
\end{center}
\vspace*{-0.5cm}
\caption{Normalized differential cross sections of $e^+e^-\to\pi^{\pm}+X$ and $e^+e^-\to K^{\pm}+X$. The points with error bars are the measured
values, where the uncertainties are the quadrature sum of the corresponding statistical and systematic uncertainties.
The $\pi^0$ and $K_S^0$ results at the same center-of-mass energies from previous measurements at BESIII~\cite{BESIII:2022zit} are also shown.
The green and red bands denote the ``NPC'' NNLO calculations~\cite{Gao:2024nkz,Gao:2024dbv} with 1$\sigma$ limits, based on new global analyses by including SIA world data~\cite{Gao:2024dbv} and our new results.
The blue band is the ``NPC'' NNLO calculation for $K_S^0$ by using $K^{\pm}$ fragmentation functions via isospin symmetry.
Only the measurements on the right of the dotted lines are employed in the fit.}
\label{final_results}
\end{figure*}
To estimate the systematic uncertainty arising from the PID efficiency matrix, a MC sampling method is applied. 
Since electrons and muons could also be produced in the decay of some unstable intermediate hadrons, their contributions to the raw counts through misidentification should be considered.
To evaluate this contribution, we first employ the same angle match method, as used in the extraction of $\bar{N}^{\rm obs}_{\pi^{\pm}/K^{\pm}}$, to determine the number of observed $e^{\pm}/\mu^{\pm}$ in the {\sc luarlw} MC sample.
Then, the PID requirements are applied on the observed $e^{\pm}/\mu^{\pm}$ to determine their contributions to  $N_{\pi^{\pm}/K^{\pm}}^{\rm raw}$.
After normalization based on the luminosity of the data, these contributions are subtracted from the nominal $N_{\pi^{\pm}/K^{\pm}}^{\rm raw}$, and the resulting relative deviations in $N_{\pi^{\pm}/K^{\pm}}^{\rm obs}$ are taken as the systematic uncertainties.
The uncertainty due to the imperfect simulation of various kinematic variables for the signal events is estimated by independently varying each selection criterion, increasing or decreasing it by 1 standard deviation of its resolution from the nominal value.
The maximum changes of the normalized differential cross sections are regarded as the systematic uncertainties.
All these individual systematic uncertainties are regarded as uncorrelated with each other and therefore are summed in quadrature.
The numbers are summarized in Supplemental Material~\cite{SuppMat}.


The normalized differential cross section for the inclusive $\pi^{\pm}$ and $K^{\pm}$ production in $e^+e^-$ annihilation at eight c.m.~energies are shown in Fig.~\ref{final_results_all} and tabulated in Supplemental Material~\cite{SuppMat}.
A new global data fit is performed for $\pi^{\pm}$ and $K^{\pm}$ FFs at next-to-next-to-leading order (NNLO) under the Nonperturbative Physics Collaboration (NPC) framework~\cite{Gao:2024nkz,Gao:2024dbv}, by incorporating existing SIA world data~\cite{Gao:2024dbv} and our new results.
The same parametrizations for FFs as in Ref.~\cite{Gao:2024dbv} are adopted.
We have assumed charge conjugation symmetry and flavor symmetries among favored (unfavored) quark FFs as in Ref.~\cite{Gao:2024dbv}.
Taking $\pi^+$ FFs as examples, we have assumed the charge conjugation symmetry
$D_q^{\pi^+}(z,Q) = D_{\bar{q}}^{\pi^-}(z, Q)$
for all scale $Q$, and flavor symmetries at the starting scale $Q_0$ among favored quark FFs
$D_u^{\pi^+}(z, Q_0) = D_{\bar{d}}^{\pi^+}(z, Q_0)$
and unfavored quark FFs
$D_{\bar{u}}^{\pi^+}(z, Q_0) = D_{{d}}^{\pi^+}(z, Q_0), D_s^{\pi^+}(z, Q_0) = D_{\bar{s}}^{\pi^+}(z, Q_0)$.
Because of the limitation of pure SIA data, we have applied an additional constraint that the $s$ quark FF shares the same shape as the $\bar{u}$ quark FF at the starting scale $Q_0$.
This results in a total of 26 and 24 free parameters for $\pi^{\pm}$ and $K^{\pm}$ FFs, respectively.
To ensure the validity of factorization and perturbative QCD (pQCD) calculations, only the results satisfying $\sqrt{s}>3$~GeV and $E_h>0.8$~GeV for BESIII measurements are employed in the fit.
Figure~\ref{final_results} illustrates the comparison between the experimental results and the fit at 3.050 and 3.671~GeV.
The overall $\chi^2/N_{pt}$ values are $294.5/365=0.81$ and $230.5/343=0.67$ for
$\pi^{\pm}$, $K^{\pm}$, respectively,
while those for BESIII $\pi^{\pm}$ and $K^{\pm}$ data are $73.4/76=0.97$ and $67.2/76=0.88$, respectively.
As shown by the green and red bands for $\pi^{\pm}$ and $K^{\pm}$, respectively,
the fitting results, based on the newly extracted FFs, can reasonably well explain the data in regions where BESIII data are included in the global analysis.
This indicates the validity of QCD factorization and fixed-order pQCD calculations at energy scales $\sqrt{s}$ down to 3~GeV.
In the low-momentum region, however, the data exhibit discrepancies compared to theoretical predictions from fixed-order pQCD calculations. This discrepancy arises primarily due to small $z$-logarithmic enhancements and higher-twist contributions, which become increasingly relevant in this regime~\cite{deFlorian:2007aj,Li:2024etc} and are not included in the NPC fits based on NNLO calculations.
Our results in this region provide valuable new inputs for advancing theoretical investigations in QCD.

The combination of the new $\pi^\pm$, $K^\pm$ and the previous $\pi^0$, $K^0_S$ measurements at BESIII~\cite{BESIII:2022zit} allows for a test of the isospin symmetry.
The measurements of the neutral pion and kaon were conducted independently, strategically taking advantage of $\pi^{0} \to \gamma \gamma$ and $K^0_S \to \pi^{+} \pi^{-}$ decay processes, thereby avoiding misidentification of signals.
As shown in Fig.~\ref{final_results},
the measured $\pi^{\pm}$ cross sections are consistent with those of $\pi^{0}$,
indicating isospin symmetry in the hadronization process of pion production,
which can be expressed in terms of the FFs as $D_i^{\pi^0} = \frac{1}{2}(D_{i}^{\pi^+} + D_{i}^{\pi^-})$ for any parton $i$.
However, the measured $K^{\pm}$ cross sections are systematically higher than those of $K^0_S$.
%
To understand the yield difference between $K^{\pm}$ and $K_S^0$,
we start with the new $K^\pm$ FFs extraction mentioned above,
from which we form a new set of $K^0_S$ FFs via ${\rm SU}(2)$ isospin symmetry.
The isospin symmetry of $K_S^0$ and $K^\pm$ FFs
has been used in global analyses of kaon FFs~\cite{deFlorian:2007aj,Albino:2008fy},
and reads
$D_q^{K_S^0} = \frac{1}{2}(D_{q'}^{K^+} + D_{q'}^{K^-})$,
with $q'=u,d$ if $q=d,u$, otherwise $q=q'$.
%
Utilizing the derived $K^0_S$ FFs,
theoretical predictions on the $K^0_S$ yield are made, as illustrated by the blue band in Fig.~\ref{final_results},
and show consistency with the BESIII $K^0_S$ results in kinematic regions where the $K^\pm$ FFs fit is performed.
%
This provides the first support for isospin symmetry at energy scales below 10~GeV in $\pi^{\pm}/\pi^0$ and $K^{\pm}/K^0_S$ fragmentation processes.
The yield difference between $K^{\pm}$ and $K_S^0$ is mainly due to
the exchange of $u$ and $d$ quark FFs while applying the isospin symmetry,
in contrast with the pion case.

In summary, we have measured the normalized differential cross sections of the $e^+e^-\to \pi^{\pm}/K^{\pm}+X$ processes using data samples collected from $\sqrt{s}=2.000$ to 3.671~GeV, with $z$ coverage from 0.13 to 0.95 and from 0.30 to 0.95 for $\pi^\pm$ and $K^{\pm}$, respectively. The data precision can reach around 1(2)\% at $z\sim0.3-0.5$ for $\pi^{\pm}$ ($K^{\pm}$).
The QCD-based analyses at NNLO under the NPC framework show that, in particular momentum regions, the data can be described reasonably well by pQCD calculations at energy scales down to 3~GeV using extracted FFs.
The $\pi^{\pm}$ yield aligns well with a set of independent $\pi^{0}$ measurements at BESIII~\cite{BESIII:2022zit}. At $E_h>0.8$~GeV, although a higher $K^{\pm}$ production cross section is observed compared to that of $K_S^0$ production, our study supports the isospin symmetry in parton fragmentation processes.
Our results fill a particular energy region where sparse inclusive charged hadron SIA data have been reported before. They provide new ingredients for future FF global data fits, thereby enhancing our understanding of hadronization in the relatively low-energy region.


The BESIII Collaboration thanks the staff of BEPCII (https://cstr.cn/31109.02.BEPC) and the IHEP computing center for their strong support. This work is supported in part by National Key R\&D Program of China under Contracts Nos. 2023YFA1606000, 2023YFA1606704, 2023YFA1609400; National Natural Science Foundation of China (NSFC) under Contracts Nos. 11635010, 11735014, 11935015, 11935016, 11935018, 12025502, 12035009, 12035013, 12061131003, 12192260, 12192261, 12192262, 12192263, 12192264, 12192265, 12221005, 12225509, 12235017, 12361141819, 12475139, 12205255, 12275173; the Chinese Academy of Sciences (CAS) Large-Scale Scientific Facility Program; the CAS Center for Excellence in Particle Physics (CCEPP); the Southern Center for Nuclear-Science Theory (SCNT); CAS Project for Young Scientists in Basic Research under Contract No. YSBR-117; Joint Large-Scale Scientific Facility Funds of the NSFC and CAS under Contract No. U1832207, No. U1732263, No. U2032105; CAS under Contract No. YSBR-101; 100 Talents Program of CAS; Postdoctoral Fellowship Program and China Postdoctoral Science Foundation under Grant Number GZC20252206; Natural Science Foundation of Gansu Province under Contract No. 23JRRA578; The Institute of Nuclear and Particle Physics (INPAC) and Shanghai Key Laboratory for Particle Physics and Cosmology; Agencia Nacional de Investigación y Desarrollo de Chile (ANID), Chile under Contract No. ANID PIA/APOYO AFB230003; German Research Foundation DFG under Contract No. FOR5327; Istituto Nazionale di Fisica Nucleare, Italy; Knut and Alice Wallenberg Foundation under Contracts Nos. 2021.0174, 2021.0299; Ministry of Development of Turkey under Contract No. DPT2006K-120470; National Research Foundation of Korea under Contract No. NRF-2022R1A2C1092335; National Science and Technology fund of Mongolia; National Science Research and Innovation Fund (NSRF) via the Program Management Unit for Human Resources \& Institutional Development, Research and Innovation of Thailand under Contract No. B50G670107; Polish National Science Centre under Contract No. 2019/35/O/ST2/02907; Swedish Research Council under Contract No. 2019.04595; The Swedish Foundation for International Cooperation in Research and Higher Education under Contract No. CH2018-7756; U. S. Department of Energy under Contract No. DE-FG02-05ER41374.

\bibliography{reference}

\begin{thebibliography}{34}
\bibitem [{bha()}]{bhabhabkgs}%
  \BibitemOpen
  \bibfield  {title} {\bibinfo {title} {{The $\cos\theta$ values of these bins
  are around $\pm0.83$, which correspond to the two gaps between barrel and
  end-cap EMC. In these bins, the Bhabha events survive the inclusive hadronic
  event selection criteria since they do not leave shower in EMC. Given that
  the fraction of these Bhabha events are less than 10\% and they are
  reasonably simulated and subtracted, we do not further suppress them in this
  analysis}},\ }\href@noop {} {\ }\BibitemShut {NoStop}%
\end{thebibliography}

\end{document}